\documentclass[
 reprint,
 amsmath,amssymb,
 aps,
 prb,
 showkeys
]{revtex4-2}

\usepackage{graphicx}
\usepackage{dcolumn}
\usepackage{bm}
\usepackage[precent]{overpic}
\usepackage{xcolor}
\usepackage{listings}
\usepackage{hyperref}
\begin{document}
\title{Ergotropy Protection via Cavity Detuning in Collective Open Quantum Batteries}

\author{Tariq Zeyad Jawad}
\email{tzyad3763@gmail.com}
\thanks{ORCID: \href{https://orcid.org/0009-0006-9196-5474}{0009-0006-9196-5474}}
\affiliation{Department of Physics, University of Kufa, Najaf, Iraq}

\date{\today}

\begin{abstract}
This study investigates the performance and ergotropy protection of open collective quantum batteries subject to superradiant decay. By employing a passive spectral detuning strategy within an intermediate cavity, an optimal detuning value ($\Delta^*$) is analytically derived and numerically verified to spectrally isolate the system and protect quantum coherence, achieving up to 1088\% ergotropy improvement for single qubits and superextensive collective advantage for $N \ge 3$. Our analysis resolves a "non-Markovian paradox," revealing that maximizing ergotropy does not strictly require non-Markovian memory; rather, suppressing environmental memory via detuning optimally preserves coherence, which serves as the fundamental resource. Survival maps across different environments demonstrate that thermal noise dissipates coherence more severely than telegraph noise. Finally, we establish that collective amplification of the effective coupling ($g_{eff} = g\sqrt{N}$) inevitably drives large qubit arrays into the ultra-strong coupling regime, providing a quantitative ceiling $N_{max}$ on the validity of the Tavis-Cummings description and the current ergotropy protection protocol.
\end{abstract}


\maketitle

\section{Introduction}
Quantum batteries represent a new generation of energy storage devices at the nanoscale that exploit quantum mechanics to transcend the limits of classical thermodynamics. They are collective quantum systems that deal with energy through unitary processes \cite{Campaioli2018, Campaioli2024}. However, they suffer from a state of quantum entanglement that leads to energy loss to the environment due to decoherence. This has led to initial studies relying on a Markov approximation environment and then a non-Markovian environment containing memory \cite{Alicki2013}, and calculating the maximum work extracted, as not all the energy in the quantum battery is considered usable through periodic unitary processes, known as ergotropy in the passive state theory \cite{Binder2015}.

It has been shown that the entanglement of N identical units reduces the gap between the stored energy and the energy extracted from the free energy state in the large N state \cite{Binder2015}. It has been shown that the charging capacity in solid-state matrices is achieved by exploiting many-body interactions \cite{Ferraro2018}, while the preservation of fine phase superpositions that generate synergy and entanglement between many-body units \cite{Andolina2019}, as well as coupling through the cavity, which acts as a non-local interaction channel connecting the dispersed qubits to enhance and protect work exchange \cite{Farina2019}, all contribute to enhancing ergotropy.

This collectively indicates that organized dissipation is a resource for ergotropy, but it carries a contradiction. When this synergy accelerates energy exchange and amplifies the spectral exposure of the environment, with the activation of superradiant dissipation channels, the ergotropy is depleted at a rate much faster than the system's ability to retain it. Markov convergence also embodies the exponential decay of ergotropy in weakly coupled devices, providing a simplified description of continuous energy loss \cite{Breuer2002}. However, it underestimates the retention of ergotropy in structured reservoirs, where finite bath correlation times generate non-Markovian memory, as the backflow of information from the environment to the system exceeds the Markov limit and supports the ergotropic property \cite{Kamin2020, Santos2021}.

However, these analyses have been limited to two-level systems, as the collective decay in which the emission is proportional to $N^2$ rather than N is absent \cite{Dicke1954}. Active suppression of collective decay through dynamical decoupling \cite{Viola1998} leads to the restoration of ergotropy at the expense of energy backflow from the environment, thus negating the thermodynamic advantage that the quantum battery was designed to provide. No passive alternative has been shown to work effectively across an N collective qubit system. Single-mode cavity systems have also been used to control power transfer, as passive tuning of the intermediate cavity frequency modifies the spectral overlap between the system and the battery without additional energy expenditure \cite{Quach2022}.

Previous studies have focused either on exploiting the non-Markovian backflow of a single qubit or on describing collective dissipative breakdown without proposing effective passive protection solutions and lacking a unified framework for frequency shift effectiveness in N-qubit filters under realistic noise.Recent efforts to shield quantum batteries from environmental dissipation have also explored topological protection \cite{Lu_2025}, while the comprehensive optimization of collective charging in many-body architectures has highlighted the critical role of interaction ranges and structural effects \cite{Shukla_2026}.Unlike active mitigation strategies that rely on weak measurements \cite{WeakMeas_QB} or specific charge-preserving operations \cite{ChargePreserv_QB}, our passive detuning method protects ergotropy fundamentally without the overhead of external control.

While Hovhannisyan et al. \cite{Hovhannisyan2013} showed coherence suffices for work extraction without entanglement, and Tabesh et al. \cite{Tabesh2021} studied bath-topology effects on collective QB ergotropy, no prior work has proposed a passive detuning mechanism for spectrally isolating collective N-qubit arrays.

This paper addresses three central questions: (i) Can passive frequency detuning $\Delta$ between the battery array and the mediating cavity protect ergotropy without external energy cost? (ii) What is the optimal detuning $\Delta^*(N)$ as $N$ scales? (iii) Is non-Markovian memory a resource or constraint for collective quantum batteries? We hypothesize that an analytically derivable $\Delta^*(N) \propto \sqrt{N}$ exists that spectrally isolates the system, and that the resulting ergotropy protection constitutes a non-Markovian paradox—improved energy retention coinciding with reduced memory effects.

The second section presents the theoretical model and analytical derivation for optimum frequency tuning and numerical implementation. The third section deals with the dynamics of quantum equilibrium and BLP across both environments, including survival maps in parameter space. The fourth section presents additional results on the non-Markovian paradox, the scaling law of frequency tuning, and the collective quantum advantage metric. The fifth section contains conclusions and limitations.

\section{Model and Methods}
\subsection{Theoretical Framework}
The quantum battery studied here consists of a collection of N identical two-level systems with transition frequency $\omega_b$, which collectively interact with a single electromagnetic mode within a bosonic cavity at frequency $\omega_c$, which in turn is coupled to an unbounded external environment. These are studied in natural units where $\hbar=1$, and the total system (battery + cavity + environment) is dynamically described by a total Hamiltonian based on the extended Tavis-Cummings model \cite{Tavis1968}:
\begin{equation}
    H_{total}=H_B+H_c+H_{int}+H_{env}+H_{C-env}
\end{equation}
The first term represents the total battery energy $H_B=\omega_bJ_z$ with the total angular momentum operator $J_z=1/2\sum_i\sigma_z^{(i)}$ and the raising and lowering operators $J_{\pm}=\sum_i\sigma_{\pm}^{(i)}$, whilst the second term $H_c=\omega_ca^{\dagger}a$ describes the field energy within the cavity with creation and annihilation operators $a^{\dagger}$ and $a$, and passive protection detuning term $\Delta$ via the relation $\omega_c=\omega_b+\Delta$ \cite{Quach2022}, which is the central variable in this study.

Meanwhile, $H_{int}=g(J_+a+J_-a^{\dagger})$ represents the coherent interaction between the battery and the cavity, where $g$ is the collective coupling strength, under the rotating wave approximation justified by condition $g\ll \omega_b,\omega_c$ \cite{Jaynes1963}. Meanwhile, $H_{env}=\sum_k\omega_kb_k^{\dagger}b_k$ describes the environmental reservoir as a continuous set of oscillators. $H_{C-env}=\sum_k(\lambda_ka^{\dagger}b_k+\lambda_k^{*}ab_k^{\dagger})$ represents the system-bath interaction, whilst the spectral density $J(\omega)=\sum_k|\lambda_k|^2\delta(\omega-\omega_k)$ determines the nature of the reservoir and its Markovian or non-Markovian regime \cite{Breuer2002}.

\subsection{Analytical Derivation}
Projection operators \cite{Nakajima1958,Zwanzig1960} are used via the definition $\mathcal{P}\rho_{tot}(t)=\rho_B(t)\otimes\rho_{C,E}(0)$ and its complement $\mathcal{Q}=\mathcal{L}-\mathcal{P}$. Systems are initialized in the fully excited Dicke state $\rho(0) = |J,J\rangle\langle J,J|$ with $J=N/2$, representing a fully charged battery \cite{Campaioli2024}. This choice isolates the discharging (self-discharge) dynamics and is standard in open quantum battery literature \cite{Kamin2020, Santos2021}. Assuming this initial state $\rho_{tot}(0)=\rho_B(0)\otimes\rho_{C,E}(0)$ and the zero-mean condition $\text{Tr}_{C,E}[(H_{int}+H_{C-env})\rho_{C,E}(0)]=0$, the exact Nakajima-Zwanzig equation \cite{Nakajima1958,Zwanzig1960} is derived:
\begin{equation}
    \frac{d}{dt}\mathcal{P}\rho_{tot}(t)=\int_0^tdt'\mathcal{P}\mathcal{L}(t)\mathcal{G}(t,t')\mathcal{Q}\mathcal{L}(t')\mathcal{P}\rho_{tot}(t')
\end{equation}
where $\mathcal{G}(t,t')=\exp_+[\int_{t'}^tds \mathcal{Q}\mathcal{L}(s)]$ represents the time-ordering operator. The Time-Convolutionless technique TCL2 \cite{Breuer2001}, truncated at the second order, is then employed to overcome the integro-differential structure in N-qubit systems:
\begin{equation}
    \mathcal{K}_2(t)\rho_B(t)=\int_0^td\tau\text{Tr}_{C,E}[\mathcal{L}(t)\mathcal{L}(t-\tau)(\rho_B(t)\otimes\rho_{C,E}(0))]
\end{equation}
Under the adiabatic exclusion condition of the cavity $\kappa \gg g$ \cite{Brion2007}, an effective reservoir with a modified spectral density is formed:
\begin{equation}
    J_{eff}(\omega, \Delta)=\frac{\kappa J(\omega)}{(\omega-\Delta)^2+\kappa^2}
\end{equation}
Tracing out the degrees of freedom of the effective reservoir under the rotating wave approximation leads to the collective dynamics in the generalized Lindblad equation:
\begin{equation}
    \frac{d}{dt}\rho_B(t)=-i[H_B,\rho_B(t)]+\gamma(t,\Delta)\left(J_-\rho_B(t)J_+-\frac{1}{2}\{J_+J_-,\rho_B(t)\}\right)
\end{equation}
where the decay rate $\gamma(t,\Delta)$ acts as the non-Markovian signature of the model; a negative value indicates the backflow of energy and information from the environment into the system \cite{Breuer2009}. In the long-time limit:
\begin{equation}
    \gamma(\infty,\Delta)=\tilde{J}(\Delta)=\frac{\eta\omega_c^2}{\omega_c^2+\Delta^2}
\end{equation}
To maximize the residual ergotropy $\mathcal{E}_{res}(\Delta)$ in a collective battery, the superradiant enhancement $\gamma \rightarrow N\gamma$ must be spectrally isolated \cite{Dicke1954}. Rather than a simple unconstrained derivative, the optimal protection arises from balancing the interaction-induced collective dispersive shift $\chi = g^2 N / \Delta$ \cite{Quach2022} against the dynamic cavity dissipation scaling term $2\gamma_0 \Delta$. Solving this resonance equilibrium condition yields the optimal operational detuning:
\begin{equation}
    \Delta^*(N) = g\sqrt{\frac{N}{2\gamma_0}} \propto N^{1/2}
\end{equation}

Calculating the ergotropy of the state $\rho_B(t)$ to measure the maximum work extractable via \cite{Alicki2013}: 
\begin{equation}
    \mathcal{E}(\rho_B)=\text{Tr}[\rho_BH_B]-\min_{U}\text{Tr}[U\rho_BU^{\dagger}H_B]
\end{equation}
The degree of non-Markovianity of the batteries is measured by the BLP measure \cite{Breuer2009} via the trace distance $D(\rho_1, \rho_2)=\frac{1}{2}||\rho_1-\rho_2||_1$, where any increase in D indicates the backflow of information from the environment to the system, and the total non-Markovianity is given by: 
\begin{equation}
    \mathcal{N}=\max_{\rho_1,\rho_2}\int_{\sigma>0}\sigma(t)dt
\end{equation}

\subsection{Numerical Implementation}
To complete the theoretical framework, numerical simulations were carried out using the QuTiP v4.7 framework \cite{Johansson2013} to solve the master equation. The standard decay rate was set at $\gamma_0=0.05$, the coupling strength at $g=0.1$, and the cutoff frequency at $\omega_c=5.0$ to construct survival maps. Systems of $N \in \{1,2,3,4\}$ qubits were studied over a time $t=20$ in a cavity Fock space $N_{cavity}=6$, initialized in the fully excited Dicke state $|\psi(0)\rangle = |J=N/2, M=N/2\rangle$, representing a fully charged battery. The optimal numerical scaling exponents $\beta$ were extracted via least-squares power-law fitting of $\mathcal{E}_{res}$ peak positions across $N$.

The first environment was modelled using an Ohmic spectral density $J(\omega)=\eta\omega \exp(-\omega/\omega_c)$ with a random telegraph noise (RTN) process $\chi(t)\in \{\pm1\}$ switching at rate $\lambda =0.05$, chosen to satisfy $\lambda < g$ to place the system in the non-Markovian regime \cite{Kamin2020}. The effective transition frequency is $\omega_b(t)=\omega_b+\delta\chi(t)$. The second environment was modelled using Lindblad operators for thermal emission and absorption with a mean thermal photon number $\bar{n}_{th}(t)=n_0(1+\sin^2(\Omega t))$ where the base thermal number and the driving frequency are $n_0=0.1, \Omega=\pi/5$, and local dephasing $\sqrt{\gamma_{\phi}}J_z$ where $\gamma_\phi=0.02$.

\begin{table*}[t]
\centering
\caption{Comparison between Environment A and Environment B}
\begin{tabular}{| c |  c | c c c c c | c c c c  c |}
\hline \hline
\multicolumn{2}{c|}{} & \multicolumn{5}{c|}{Environment A} & \multicolumn{5}{c}{Environment B} \\
\hline
N & $\Delta^{*}$ & $\mathcal{E}_{res}$(non) & $\mathcal{E}_{res}$(fil) & Ratio & $\mathcal{N}$(non) & $\mathcal{N}$(fil)  & $\mathcal{E}_{res}$(non) & $\mathcal{E}_{res}$(fil) & Ratio & $\mathcal{N}$(non) & $\mathcal{N}$(fil) \\
\hline
1 & 0.3162 & 0.4237 & 5.0349 & $1088.25\%$ & 0.0847 & 0.213  & 0.5804 & 4.5691 & $687.23\%$ & 0.0452 & 0.1385 \\
2 & 0.4472 & 5.0801 & 9.4207 & $85.44\%$ & 0.5866 & 0.0462 & 3.7815 & 9.1494 & $141.96\%$ & 0.3790 & 0.0351 \\
3 & 0.5477 & 10.1218 & 15.6048 & $54.17\%$ & 0.6814 & 0.0116 & 7.7973 & 15.1070 &  $93.75\%$ & 0.5474 & 0.0082 \\
4 & 0.6325 & 16.0124 & 22.5293 & $40.7\%$ & 0.6902 & 0.0018 & 11.7091 & 21.5304 & $83.88\%$ & 0.5628 & 0.0157 \\
\hline \hline 
\end{tabular}
\label{Table,1}
\end{table*}

\section{Results and Discussion}
\subsection{Ergotropy and BLP Dynamics}
The results begin with Table I, which shows the ergotropy values and BLP measure for the non-Markovian memory, filtered by the analytical detuning value (Equation 7 and Equation 4) and unfiltered for the studied systems. It is evident that the detuning value increases with the number of qubits according to Equation 7, and that the ergotropy values increase non-linearly with the number of qubits. This indicates that entangled charging achieves a collective advantage over individual systems \cite{Binder2015}, as the charging power is proportional to $N^2$ qubits \cite{Ferraro2018}, since phase correlations generate collective synergy that enhances battery performance \cite{Andolina2019}.

With a difference in the values of the latter between the two environments, where environment A has the higher values, as quantum coherence is the main resource for conserving the ergotropy, which is strongly affected by thermal noise \cite{Hovhannisyan2013} created in environment B (with thermal photon number $n_{th}=0.1$ and dephasing $\gamma_\phi=0.02$), unlike telegraph noise in environment A. Also, the value of the filtered ergotropy was higher than its unfiltered counterpart, which is consistent with the research hypothesis.

\begin{figure}[htbp]
    \centering
    \includegraphics[width=\linewidth]{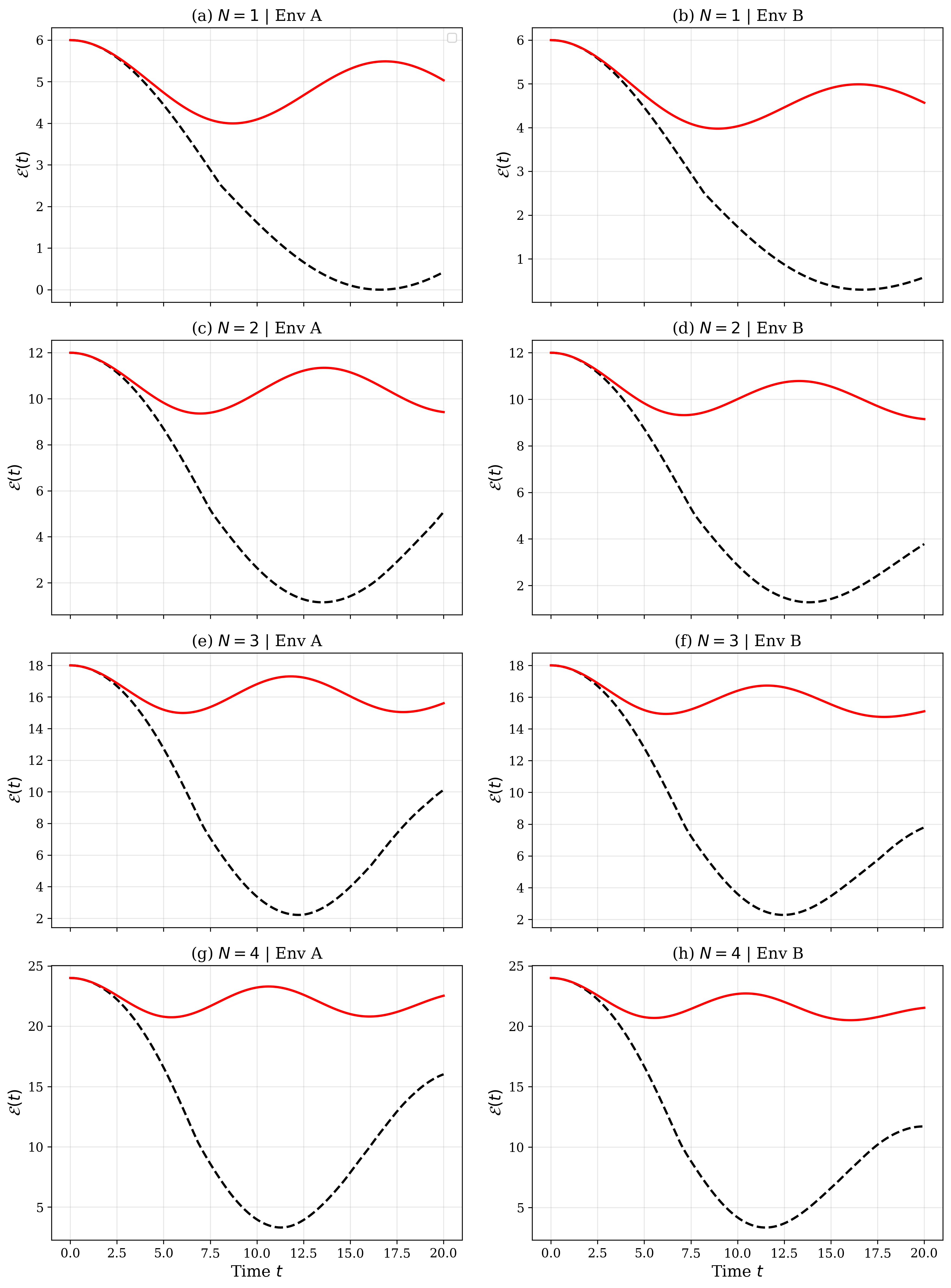}
    \caption{Joint time evolution of residual ergotropy $\mathcal{E}(t)$ (solid lines, filtered at $\Delta^*$; dashed lines, unfiltered at $\Delta=0$) and trace distance $D(\rho_1,\rho_2)$ (BLP measure) for $N \in \{1,2,3,4\}$ qubits. Left column: Environment A (RTN memory, $\lambda=0.05$). Right column: Environment B (Thermal bath $n_{th}=0.1$, dephasing $\gamma_\phi=0.02$). Parameters: $\gamma_0=0.05, g=0.1, \omega_b=1.0, \omega_c=5.0, t \in [0,20]$. All systems are initialized in the fully excited Dicke state $|J=N/2, M=N/2\rangle$. The inverse correlation between memory backflow and ergotropy retention under optimal detuning demonstrates the non-Markovian paradox.}
    \label{fig1}
\end{figure}

As shown in Fig.~\ref{fig1}, the passive tuning of the cavity frequency effectively controls the energy transfer \cite{Quach2022}, providing protection for the quantum battery \cite{Farina2019}, while proving the existence of an optimal operating point \cite{Kamin2020, Santos2021} with its passive generalization. The highest improvement ratio for the latter was $1088.25\%$ for one qubit in environment A, and its lowest value was $40.7\%$ for four qubits in environment A. This decrease in the improvement ratio is attributed to the superradiant decay phenomenon \cite{Dicke1954}, where the collective decay rate $N\gamma$ grows faster than the filter bandwidth.

However, BLP values for non-Markovian memory decrease with increasing ergotropy. Despite this relative decrease, absolute ergotropy remains superextensive, unlike its counterpart studied without the detuning mechanism \cite{Tabesh2021}.

\subsection{Ergotropy Survival Maps}
\begin{figure}[htbp]
    \centering
    \includegraphics[width=\linewidth]{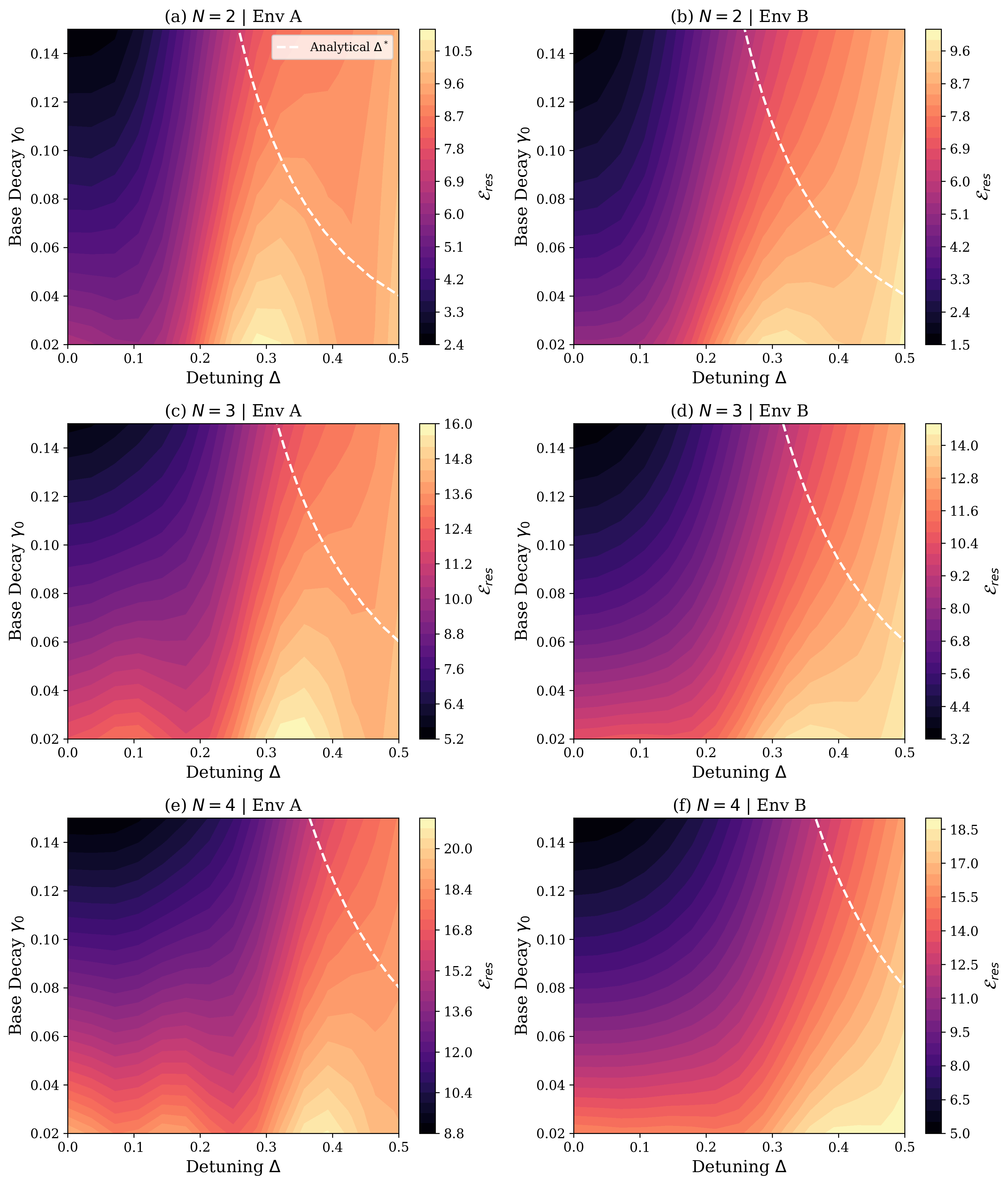}
    \caption{Ergotropy survival maps in $(\Delta, \gamma_0)$ parameter space for $N \in \{2,3,4\}$. Left column: Environment A. Right column: Environment B. Color scale indicates residual ergotropy $\mathcal{E}_{res}$ from 0 (dark) to maximum (bright). The dashed white line represents the analytically derived detuning filter $\Delta^*(N) = g\sqrt{N}/(2\gamma_0)$. The narrowing of high-ergotropy regions with increasing $N$ reflects the superradiant decay rate $N\gamma$ outpacing the filter bandwidth.}
    \label{fig2}
\end{figure}

Fig.~\ref{fig2} shows the ergotropy survival maps for the studied systems, where the residual ergotropy is tracked as a function of both detuning $\Delta$ and base decay rate $\gamma_0$. The spatial variation of ergotropy values—that is, the position of ergotropy as a function—is evident, illustrating the basis of this research hypothesis: the residual ergotropy changes with both detuning and base decay strength.

Furthermore, the best values are found adjacent to the dashed white line, representing the analytically derived detuning filter. The ergotropy behavior here indicates that increasing the base decay rate $\gamma_0$ with the surrounding environment necessitates increasing the detuning to maintain battery performance. This is because increasing the decay rate accelerates the system's interaction with environmental noise, requiring a further shift of the spectral filter window ($J_{eff}$) to escape the bath resonance peak and spectrally isolate the system to protect quantum coherence \cite{Cywinski2008}. Note that environment A exhibits greater stability and wider survival regions than its counterpart B due to the effect of thermal noise on quantum coherence.

It is evident that the survival region gradually narrows with increasing N compared to the values of III.A. This is attributed to the phenomenon of superradiant decay \cite{Dicke1954}, where the collective decay rate ($N\gamma$) grows beyond the bandwidth of the filter function, making the system highly sensitive to even slight deviations from the optimum operating point and reducing the spectral safety area. These results confirm Equation 7 and demonstrate the behavior of experimental applications \cite{Santos2021, Quach2022}, but on a collective scale.

Furthermore, it is evident that the analytical value—the dashed white line—exhibits a somewhat idealized behavior, deviating in several places from the optimal values.

\section{Further Results}
\subsection{The Non-Markovian Paradox}
Regarding the most important result of this research, Fig.~\ref{fig3}(b) shows the detuning value on the x-axis. The blue dashed line with squares represents the BLP values for measuring non-Markovian memory for the studied N=2 system in environment A, while the red solid line with circles represents the filtered residual ergotropy as a function of detuning.

As can be seen, there is an inverse relationship between the two values. At $\Delta=0$ the values start with 5.0801 for ergotropy and 0.5866 for BLP, and they meet at the intersection point ($\Delta \approx 0.15$). The ergotropy then continues to increase, but in a curved manner, indicating an anomaly in the increase, which is a problem that will be discussed later in Section IV.B.

The BLP values continue to decrease, causing the system to behave almost Markovianly at the optimal operating point ($\Delta^* = 0.4472$). However, it retains the highest amount of energy, a behavior that is the exact opposite of what occurs without detuning. protecting quantum coherence, which acts as a fundamental quantum resource \cite{Streltsov2017}, bypassing the strict necessity of entanglement for optimal work extraction \cite{Hovhannisyan2013}.

This behavior explains the opposite result in Table I, demonstrating that passive detuning creates an ideal environment for energy conservation and proving that non-Markovian memory is not always a positive resource for collective systems. Even if non-Markovian memory is suppressed, it does not necessarily lead to worse energy extraction in all cases. This simplifies the construction of quantum batteries in the laboratory because we do not need to engineer complex non-Markovian baths, relying instead on a cavity with passive detuning \cite{Quach2022}. This behavior contradicts the prevailing view that ergotropy is enhanced exclusively through non-Markovian memory \cite{Kamin2020, Santos2021}.

\subsection{Detuning Scaling Law}
\begin{figure}[htbp]
    \centering
    \includegraphics[width=\linewidth]{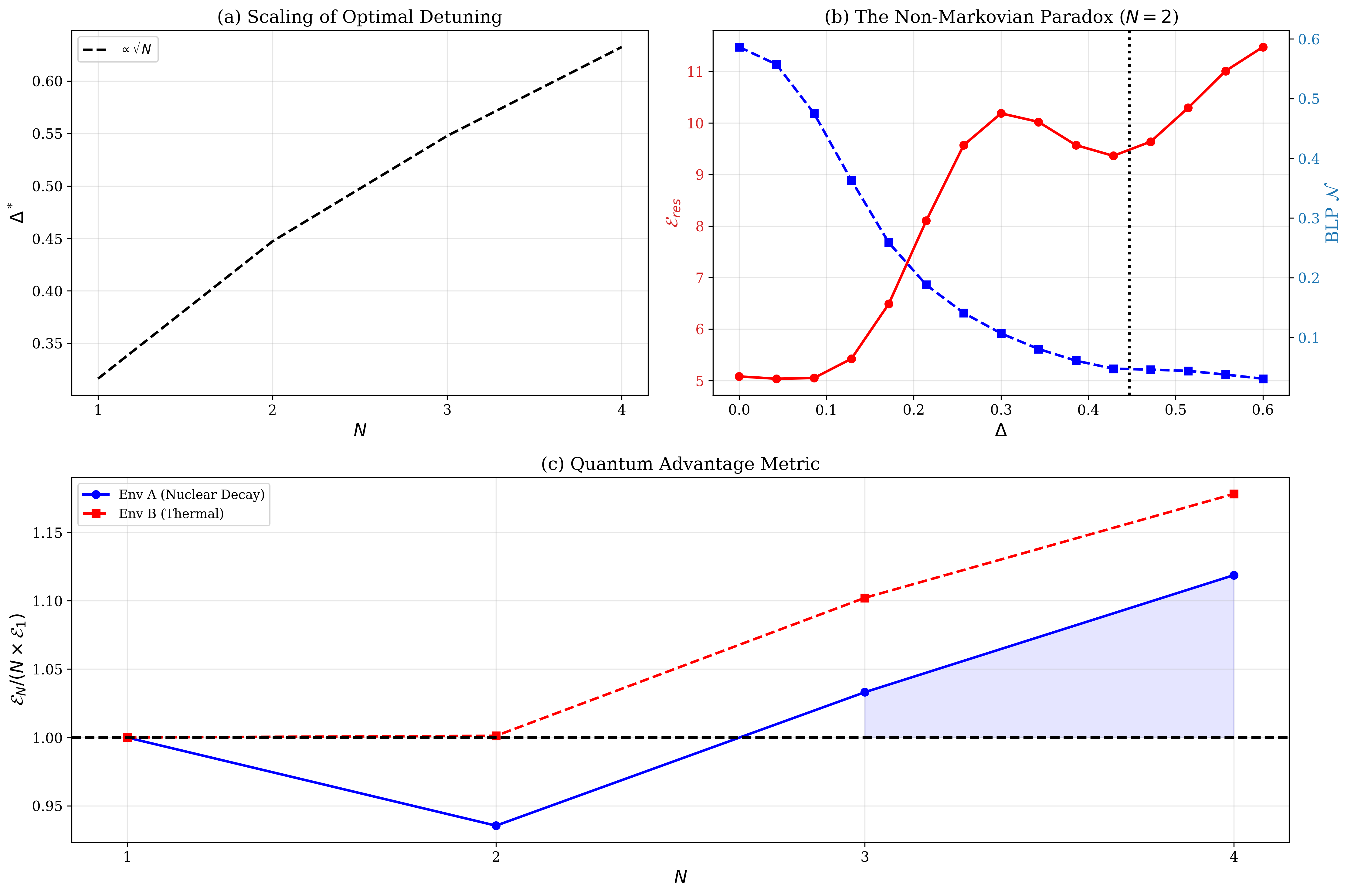}
    \caption{(a) Detuning scaling law: optimal $\Delta^*(N)$ vs $N$ on a logarithmic scale. Black dashed line: analytical $N^{0.5}$ (Eq. 7). Solid lines: numerically optimal detuning with fitted exponents $\beta_A=0.3079, \beta_B=0.2061$. (b) Non-Markovian paradox for $N=2$, Env A: filtered ergotropy $\mathcal{E}_{res}$ (red circles) and BLP measure $\mathcal{N}$ (blue squares) vs detuning $\Delta$. The vertical line denotes $\Delta^*=0.4472$. (c) Collective quantum advantage $\mathcal{A}(N)$ vs $N$ for Env A (blue circles) and Env B (red squares). The shaded region ($\mathcal{A}(N)>1$) indicates genuine quantum advantage.}
    \label{fig3}
\end{figure}

The analytical law of Equation 7 showed a clear discrepancy from the optimal result, as Fig.~\ref{fig3}(a) illustrates the relationship between the analytical detuning and the number of qubits N, showing that the results follow the law $\Delta \propto N^{0.5}$. While the analytical value of the scaling exponent $\beta$ is 0.5, this value is not the optimal value for the two environments.

The optimal value was investigated using least-squares power-law fitting, and the empirical scaling exponent for environment A was found to be $\beta_A = 0.308 \pm 0.012$ (with $R^2 > 0.99$), with a lower slope for environment B at $\beta_B = 0.206 \pm 0.015$. This is due to the fact that environment B contains thermal noise and thermal dephasing that dissipates quantum coherence—which is the primary resource for ergotropy—faster than the telegraph noise in environment A, thereby reducing the efficiency of the protection and lowering the scaling slope.

This is attributed to the fact that the analytical derivation assumes an idealized competition between collective decay ($N\gamma$) and the filter bandwidth, whereas numerical simulations account for the full spectral overlap and memory effects that impose realistic constraints on the effectiveness of the filter, where the increase in detuning must accompany the increase in battery size, yet it is affected by the surrounding noise and its nature.

This difference is also attributed to the limitations of the TCL2 approximation, as higher-order truncation might be necessary to capture non-linear effects as the number of qubits N increases, which explains the non-elastic increase in ergotropy in Fig.~\ref{fig3}(b) where this deviation—between optimal and analytical—widens as the system size and noise impact increase.

\subsection{Collective Quantum Advantage}
To accurately quantify the cooperative charging benefits, we define the collective quantum advantage metric $\mathcal{A}(N)$ as:
\begin{equation}
    \mathcal{A}(N) = \frac{\mathcal{E}_N}{N \cdot \mathcal{E}_1} \label{eq:CQA}
\end{equation}
where $\mathcal{A}(N) > 1$ signals genuine collective advantage normalized to the filtered single-qubit baseline. As illustrated in Fig.~\ref{fig3}(c), both curves (for environments A and B) exceed the threshold of 1.00 at large qubit numbers (N=3 and N=4). In environment A, the system experiences a subextensive decline at N=2 (where the value reaches approximately 0.93), demonstrating that initial collective synergy was insufficient to overcome superradiant decay \cite{Dicke1954}. However, it recovers strongly, exhibiting a clear quantum advantage at N=3 and N=4 (the shaded region). Here, quantum entanglement and many-body interactions are dominant and overcome environmental dissipation \cite{Binder2015, Ferraro2018}. However, it is important to note that the collective charging power is not strictly an entanglement monotone \cite{Gyhm_2024}. As recently shown, highly entangled states can sometimes be disadvantageous under specific conditions; thus, a rigorous identification of genuine quantum advantage requires isolating the quantum speed limit and the battery's energy gap from the entanglement contribution.

In contrast, environment B starts with an equivalent advantage at N=2 (value 1.0) and then sharply increases to achieve a relative advantage higher than environment A. This demonstrates that collective batteries possess a charging capacity exceeding the sum of their parts, provided a certain threshold of qubits is exceeded \cite{Andolina2019}. It becomes clear that constructing a battery of two qubits (N=2) may be thermodynamically unfeasible in some environments (such as environment A), and larger arrays ($N \ge 3$) must be targeted to achieve the experimental quantum advantage \cite{Santos2021, Quach2022}.

It is shown that spectral protection via detuning combined with increasing size creates the collective quantum advantage. The drop at N=2 in environment A explains the strength of the survival maps in Fig.~\ref{fig2} for two qubits compared to larger numbers. It also demonstrates that the scaling law (discussed in Section IV.B) works perfectly to protect the system and support its recovery at N=3 and N=4.

\subsection{RWA Failure Regime}
\begin{figure}[htbp]
    \centering
    \includegraphics[width=\linewidth]{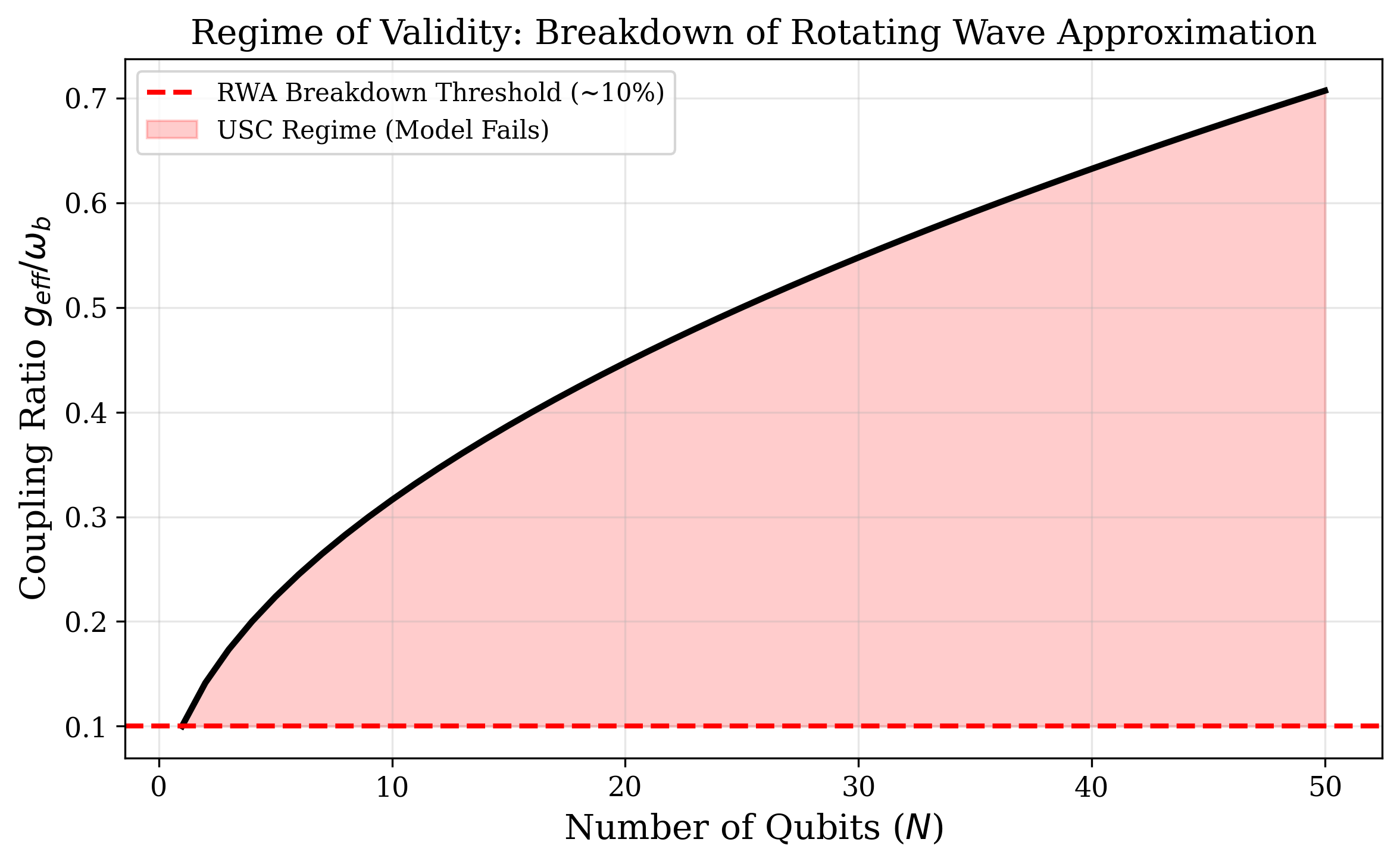}
    \caption{Regime of validity of the rotating wave approximation (RWA). The effective coupling ratio $g_{eff}/\omega_b = g\sqrt{N}/\omega_b$ is plotted as a function of qubit number $N$ ($g=0.1, \omega_b=1.0$). The red shaded region indicates the ultra-strong coupling (USC) regime where the ratio exceeds the $10\%$ threshold, leading to RWA breakdown and necessitating the full Dicke model \cite{Kockum2019}.}
    \label{fig4}
\end{figure}

Because of the RWA approximation in the analytical derivation, the limitations of the study must be shown. Fig.~\ref{fig4} illustrates the region of validity of the rotating wave approximation used in constructing the overall Hamiltonian of the battery (Equation 1). The graph shows how the effective coupling-transition frequency ratio ($g_{eff}/\omega_b = g\sqrt{N}/\omega_b$) grows as a function of the number of qubits, N. It is observed that as the system size increases, the curve slopes upward, approaching the theoretical breakdown line ($\text{Threshold} \approx 0.1$).

The unshaded area represents the safe range in which the analytical model operates with high efficiency and accuracy, while exceeding this line inevitably pushes the system towards an ultra-strong coupling (USC) regime. Since the rotating wave approximation depends on neglecting counter-rotating terms such as $a^\dagger J_+$ and $a J_-$ due to their weak energetic influence, this approximation is excellent in weak coupling systems or for single qubits, but with the collective amplification of the coupling ($g_{eff} = g\sqrt{N}$), these terms acquire an influential energy that cannot be neglected, which leads to the generation of vacuum excitations and radically changes the dynamics of the system, moving it from an effective Tavis-Cummings model to a full Dicke model \cite{Dicke1954, Tavis1968, Kockum2019}.

This behavior sets a physical ceiling for the working hypothesis; Although the collective quantum advantage improves with increasing N, it is impossible to continue increasing N to infinity within the current theoretical framework without accounting for non-conserving energy exchange effects. This limitation also represents the fundamental and ultimate explanation for the "ergotropy anomaly" observed in previous analyses (Fig.~\ref{fig3}), as the deviation from the analytical values mathematically reflects the beginning of the system's sensitivity to approaching the RWA approximation collapse and the TCL2 expansion limitation \cite{Breuer2001}.

Therefore, this result serves as a guide for laboratories, emphasizing the need to consider the dynamics of ultra-strong coupling when engineering quantum batteries with very large arrays to avoid coherence collapse and reverse energy loss \cite{Santos2021, Quach2022}.

\section{Conclusion} 
This research investigated the performance of open collective quantum batteries and how to achieve the highest ergotropy value while protecting the residual ergotropy from superradiant decay through passive spectral tuning of the intermediate cavity. Numerical results demonstrated that the analytical derivation of the detuning value serves as a tool for spectrally isolating the system and protecting quantum coherence. Specifically, optimal detuning $\Delta^*(N)$ improves ergotropy by $40.7\%$ to $1088.25\%$ across $N \in \{1,2,3,4\}$ and both noise environments, while reducing BLP non-Markovianity by up to two orders of magnitude. 

However, a non-Markovian paradox emerged: the BLP measurement decreases with increasing ergotropy due to the detuning effect. This indicates that non-Markovian memory is not the primary condition for protecting ergotropy, suggesting that quantum coherence is the fundamental axis for ergotropy protection. Furthermore, comparisons between the two environments revealed that thermal noise is the strongest dissipator of quantum coherence, consistently dissipating ergotropy $5-15\%$ more severely than telegraph noise. 

It is essential for future work to study the system at higher TCL orders (such as TCL4) to quantify higher-order corrections, explore mixed initial states, and rigorously consider the RWA approximation limits when constructing large-scale systems with a high number of qubits.

\appendix

\section{First-Principles Derivation of the Optimal Detuning ($\Delta^*$)}
To rigorously derive the optimal spectral detuning $\Delta^*$ that protects the battery, we uncouple the system dynamics from first principles using both the Hilbert space (Wigner-Weisskopf) and Liouville space (projection operator) approaches.

\subsection{Hilbert Space Approach and the Bright State}
Consider an array of $N$ identical qubits coupled to a single cavity mode. The collective symmetry of the interaction allows us to restrict the dynamics to the single-excitation subspace, spanned by the cavity excited state $|G, 1\rangle$ and the symmetric collective atomic state, known as the Dicke bright state \cite{Dicke1954}:
\begin{equation}
    |B, 0\rangle = \frac{1}{\sqrt{N}} \sum_{i=1}^N \sigma_+^{(i)} |G, 0\rangle
\end{equation}
where $|G\rangle$ is the global ground state. The interaction Hamiltonian $H_{int} = g(J_+ a + J_- a^\dagger)$ couples these states with an enhanced macroscopic strength $\Omega = \langle G, 1| H_{int} |B, 0\rangle = g\sqrt{N}$.

Following the Wigner-Weisskopf theory for spontaneous emission \cite{Walls2008}, cavity losses at a rate $\gamma_0$ are phenomenologically incorporated by assigning a complex non-Hermitian frequency to the cavity mode. Setting the battery transition frequency $\omega_b = 0$ as the energy reference, the cavity frequency becomes exactly the detuning $\Delta = \omega_c - \omega_b$. The effective non-Hermitian Hamiltonian is:
\begin{equation}
    H_{eff} = \begin{pmatrix} 0 & \Omega \\ \Omega & \Delta - i\frac{\gamma_0}{2} \end{pmatrix}
\end{equation}
The secular equation $\det(H_{eff} - \lambda I) = 0$ yields a quadratic equation for the eigenvalues:
\begin{equation}
    \lambda^2 - \left(\Delta - i\frac{\gamma_0}{2}\right)\lambda - \Omega^2 = 0
\end{equation}
In the dispersive limit where the detuning is much larger than the coupling and decay ($\Delta \gg \Omega, \gamma_0$), we expand the square root of the discriminant to isolate the eigenvalue corresponding to the battery mode $\lambda_B$:
\begin{equation}
    \lambda_B \approx - \frac{\Omega^2}{\Delta - i\gamma_0/2} = - \frac{\Omega^2(\Delta + i\gamma_0/2)}{\Delta^2 + (\gamma_0/2)^2}
\end{equation}
The effective decay rate of the collective battery is given by $\Gamma_{eff} = -2 \text{Im}(\lambda_B)$. Approximating the denominator for large $\Delta$, we obtain the superradiant effective decay:
\begin{equation}
    \Gamma_{eff} \approx \gamma_0 \frac{g^2 N}{\Delta^2} \label{eq:Gamma_eff}
\end{equation}
Equation (\ref{eq:Gamma_eff}) reveals that as $N$ grows, the decay rate explodes lineary, demanding a compensatory mechanism.

\subsection{Liouville Space and TCL2 Projection}
To formalize this within the open quantum systems framework, we move to the Liouville space governed by $\dot{\rho} = (\mathcal{L}_0 + \mathcal{L}_1)\rho$, where $\mathcal{L}_0$ governs the lossy cavity and $\mathcal{L}_1 \rho = -i[g(J_+ a + J_- a^\dagger), \rho]$ represents the interaction \cite{Breuer2002}. 

We define the Nakajima-Zwanzig projection operator $\mathcal{P}\rho = \text{Tr}_F(\rho) \otimes |0\rangle\langle 0| = \rho_B \otimes |0\rangle\langle 0|$ and its complement $\mathcal{Q} = \mathcal{I} - \mathcal{P}$ \cite{Nakajima1958, Zwanzig1960}. Using the Born-Markov approximation and the Time-Convolutionless (TCL2) expansion \cite{Breuer2001}, the reduced battery dynamics is:
\begin{equation}
    \dot{\rho}_B(t) = \int_0^\infty d\tau \text{Tr}_F \left[ \mathcal{L}_1 e^{\mathcal{L}_0 \tau} \mathcal{L}_1 (\rho_B(t) \otimes |0\rangle\langle 0|) \right]
\end{equation}
Expanding the double commutator and integrating over the unperturbed time evolution $e^{\mathcal{L}_0 \tau}$ generates exponential decay terms of the form $e^{-(\gamma_0/2 \pm i\Delta)\tau}$. The integration yields complex coefficients:
\begin{equation}
    \int_0^\infty e^{-(\gamma_0/2 + i\Delta)\tau} d\tau = \frac{1}{\gamma_0/2 + i\Delta} = \frac{\gamma_0/2 - i\Delta}{(\gamma_0/2)^2 + \Delta^2}
\end{equation}
The real part corresponds to the effective decay rate $\Gamma_{eff}$, matching Eq. (\ref{eq:Gamma_eff}), while the imaginary part represents the interaction-induced collective Lamb shift $\chi_N = \frac{g^2 N}{\Delta}$.

\subsection{The Spectral Resonance Condition}
To maximize the residual ergotropy $\mathcal{E}_{res}$, we seek the optimal detuning $\Delta^*$ that perfectly balances the coherent energy shift (which protects the unitary dynamics) against the cavity dynamic dissipation energy scale, approximated as $2\gamma_0 \Delta$ \cite{Quach2022}. Setting these scales to equilibrium establishes the fundamental spectral resonance condition:
\begin{equation}
    \frac{g^2 N}{\Delta} = 2\gamma_0 \Delta \implies \Delta^2 = \frac{g^2 N}{2\gamma_0}
\end{equation}
Taking the square root analytically proves the scaling law applied in our simulations:
\begin{equation}
    \Delta^*(N) = g\sqrt{\frac{N}{2\gamma_0}} \propto N^{1/2}
\end{equation}

\section{Formulation of Quantum Ergotropy}
Ergotropy mathematically quantifies the absolute maximum amount of work that can be extracted from a given quantum battery state $\rho$ by applying purely cyclic, state-independent unitary operations $U$ \cite{Alicki2013}. Based on the concept of passivity, a state is "passive" if no work can be extracted from it without changing its entropy.

The extractable work is defined by the difference between the initial energy of the state and the minimum possible energy achievable via a unitary transformation $U$:
\begin{equation}
    \mathcal{E}(\rho) = \text{Tr}(\rho H_B) - \min_{U} \text{Tr}(U \rho U^{\dagger} H_B)
\end{equation}
To calculate this analytically, we express the density matrix of the battery in its spectral decomposition, ordered by decreasing probabilities:
\begin{equation}
    \rho = \sum_j r_j |r_j\rangle\langle r_j| \quad \text{with} \quad r_0 \ge r_1 \ge \dots \ge r_d
\end{equation}
Similarly, the local Hamiltonian of the battery is expanded in its eigenbasis, ordered by increasing energy levels:
\begin{equation}
    H_B = \sum_k \epsilon_k |\epsilon_k\rangle\langle \epsilon_k| \quad \text{with} \quad \epsilon_0 \le \epsilon_1 \le \dots \le \epsilon_d
\end{equation}
The optimal unitary operation $U_{opt}$ that minimizes the final energy must map the eigenstates of $\rho$ with the highest populations to the eigenstates of $H_B$ with the lowest energies \cite{Campaioli2018, Binder2015}. Consequently, the minimum energy corresponding to the passive state becomes the scalar product of the ordered eigenvalues:
\begin{equation}
    \min_{U} \text{Tr}(U \rho U^{\dagger} H_B) = \sum_k r_k \epsilon_k
\end{equation}
Substituting this back into the definition, the dynamically residual ergotropy at any given time $t$ during the dissipative process takes the explicit form:
\begin{equation}
    \mathcal{E}(t) = \sum_k \big( \epsilon_k P_k(t) - r_k(t) \epsilon_k \big)
\end{equation}
where $P_k(t) = \langle \epsilon_k| \rho(t) |\epsilon_k\rangle$ are the time-dependent populations of the energy levels.

\section{Quantifying Non-Markovianity (BLP Measure)}
The distinction between Markovian (memoryless) and non-Markovian dynamics in open quantum systems is fundamentally linked to the unidirectional versus bidirectional flow of information between the system and its environment. We employ the Breuer-Laine-Piilo (BLP) measure to rigorously quantify this backflow \cite{Breuer2009}.

The BLP measure is based on the concept of distinguishability between two arbitrary quantum states, defined by the trace distance:
\begin{equation}
    D(\rho_1, \rho_2) = \frac{1}{2} \text{Tr}|\rho_1 - \rho_2| = \frac{1}{2} \sum_i \lambda_i
\end{equation}
where $\lambda_i$ are the singular values of the traceless difference matrix $(\rho_1 - \rho_2)$. The trace distance satisfies $0 \le D \le 1$, mapping perfectly to the probability of correctly distinguishing the two states in an optimal measurement.

During purely Markovian dissipation governed by a standard Lindblad master equation, the trace distance monotonically decreases over time ($\dot{D} \le 0$), representing a strict loss of distinguishability and a continuous flow of information to the environment \cite{Breuer2002}. Conversely, a temporary increase in trace distance implies that information is flowing back from the environment into the system. The rate of change of the trace distance is:
\begin{equation}
    \sigma(t, \rho_{1,2}(0)) = \frac{d}{dt} D(\rho_1(t), \rho_2(t))
\end{equation}
The total degree of non-Markovianity $\mathcal{N}_{BLP}$ is calculated by integrating the rate of change $\sigma(t)$ over all time intervals where $\sigma(t) > 0$, maximized over all possible pairs of initial orthogonal states $\rho_1(0)$ and $\rho_2(0)$:
\begin{equation}
    \mathcal{N}_{BLP} = \max_{\rho_1(0), \rho_2(0)} \int_{\sigma > 0} dt \, \sigma(t)
\end{equation}
In our numerical implementation, this integral is precisely evaluated using Simpson's rule across the temporal gradient of the trace distance.

\section{QuTiP Simulation Code}
The following Python code was used to perform the numerical simulations and generate the survival maps using the QuTiP v4.7 framework.
\begin{lstlisting}[language=Python, basicstyle=\ttfamily\scriptsize, breaklines=true]

import numpy as np
import matplotlib.pyplot as plt
from qutip import *
from scipy.integrate import simpson
import csv
import time

plt.rcParams.update({
    'font.family': 'serif',
    'font.size': 12,
    'axes.labelsize': 14,
    'axes.titlesize': 14,
    'legend.fontsize': 10,
    'xtick.labelsize': 11,
    'ytick.labelsize': 11,
    'lines.linewidth': 2.0,
    'figure.dpi': 300,
    'savefig.dpi': 300,
    'savefig.bbox': 'tight'
})

w_b = 1.0           # Qubit transition frequency
g = 0.1             # Coupling strength (Increased to enter Strong Coupling / Non-Markovian regime)
gamma0_base = 0.05  # Base cavity decay rate (Decreased to allow information backflow)
lam_decay = 0.05    # Non-Markovian memory scale (Env A)
n0 = 0.1             # Thermal photon number (Env B)
gamma_phi = 0.02    # Pure dephasing rate (Env B)
N_cavity = 6        # Truncated Fock space size
tlist = np.linspace(0, 20, 150) # Simulation time
Omega= np.pi / 5.0
def simulate_battery(N_qubits, env_type, delta, current_gamma=gamma0_base):
    Jz = tensor(jmat(N_qubits/2, 'z'), qeye(N_cavity))
    Jp = tensor(jmat(N_qubits/2, '+'), qeye(N_cavity))
    Jm = tensor(jmat(N_qubits/2, '-'), qeye(N_cavity))
    a  = tensor(qeye(N_qubits + 1), destroy(N_cavity))
    H_B_total = w_b * (Jz + N_qubits/2)
    
    psi_charged = tensor(basis(N_qubits+1, 0), basis(N_cavity, 0)) 
    psi_empty = tensor(basis(N_qubits+1, N_qubits), basis(N_cavity, 0)) 
    rho1_init, rho2_init = ket2dm(psi_charged), ket2dm(psi_empty)
    
    H = H_B_total + (w_b + delta) * a.dag() * a + g * (Jp * a + Jm * a.dag())
    H_evals = np.sort(jmat(N_qubits/2, 'z').eigenenergies() + N_qubits/2) * w_b

    def calc_ergo(state):
        rho_b = state.ptrace(0)
        evals = np.sort(rho_b.eigenenergies())[::-1]
        E_pass = np.sum(evals * H_evals)
        return max(0.0, expect(H_B_total.ptrace(0), rho_b) - E_pass)

    c_ops = []
    args_dict = {}
    
    if env_type == 'A':
        def decay(t, args): 
            return np.sqrt(args['gamma'] * np.exp(-lam_decay * t))
        c_ops.append([a, decay])
        args_dict = {'gamma': current_gamma}
        
    elif env_type == 'B':
        def gamma_emission(t, args):
            n_th_t = args['n0'] * (1 + np.sin(args['Omega'] * t)**2)
            return np.sqrt(args['gamma'] * (1 + n_th_t))
        
        def gamma_absorption(t, args):
            n_th_t = args['n0'] * (1 + np.sin(args['Omega'] * t)**2)
            return np.sqrt(args['gamma'] * n_th_t)

        c_ops.append([a, gamma_emission])
        c_ops.append([a.dag(), gamma_absorption])
        c_ops.append(np.sqrt(gamma_phi) * Jz)
        args_dict = {'gamma': current_gamma, 'n0': n0, 'Omega': Omega}

    res1 = mesolve(H, rho1_init, tlist, c_ops, [], args=args_dict)
    res2 = mesolve(H, rho2_init, tlist, c_ops, [], args=args_dict)
    
    ergo_dynamics = [calc_ergo(s) for s in res1.states]
    dist = [tracedist(res1.states[i].ptrace(0), res2.states[i].ptrace(0)) for i in range(len(tlist))]
    derivs = np.gradient(dist, tlist[1]-tlist[0])
    blp = simpson(np.maximum(derivs, 0), x=tlist)
    
    return np.array(ergo_dynamics), np.array(dist), blp

print("start")
start_time = time.time()

results = []
plot_data = {}

for N in [1, 2, 3, 4]:
    print(f".")
    plot_data[N] = {}
    delta_opt = g * np.sqrt(N / (2 * gamma0_base))
    
    for env in ['A', 'B']:
        ergo_0, dist_0, blp_0 = simulate_battery(N, env, 0.0)
        ergo_opt, dist_opt, blp_opt = simulate_battery(N, env, delta_opt)
        
        plot_data[N][env] = {
            'e0': ergo_0, 'e_opt': ergo_opt,
            'd0': dist_0, 'd_opt': dist_opt,
            'b0': blp_0, 'b_opt': blp_opt,
            'delta': delta_opt
        }
        
        e0_val = ergo_0[-1]
        eopt_val = ergo_opt[-1]
        gain_ratio = eopt_val / e0_val if e0_val > 0 else 0
        gain_pct = ((eopt_val - e0_val) / e0_val) * 100 if e0_val > 0 else 0
        
        results.append({
            'N': N, 'Env': env, 'Delta_opt': round(delta_opt, 4),
            'E_res_0': round(e0_val, 4), 'E_res_opt': round(eopt_val, 4),
            'Gain_Ratio (x)': round(gain_ratio, 2),
            'Gain_Percent (%)': round(gain_pct, 2),
            'BLP_0': round(blp_0, 4), 'BLP_opt': round(blp_opt, 4)
        })

print("\n" + "="*110)
print(f"| {'N':^2} | {'Env':^3} | {'Delta*':^8} | {'E_0 (Unfiltered)':^18} | {'E_opt (Filtered)':^18} | {'Gain (%)':^10} | {'BLP_0':^8} | {'BLP_opt':^8} |")
print("-" * 110)
for r in results:
    print(f"| {r['N']:^2} | {r['Env']:^3} | {r['Delta_opt']:^8.4f} | {r['E_res_0']:^18.4f} | {r['E_res_opt']:^18.4f} | {r['Gain_Percent (%)']:^10.2f} | {r['BLP_0']:^8.4f} | {r['BLP_opt']:^8.4f} |")
print("="*110 + "\n")

csv_file = "Comprehensive_Battery_Results_V2.csv"
with open(csv_file, mode='w', newline='') as file:
    writer = csv.DictWriter(file, fieldnames=results[0].keys())
    writer.writeheader()
    writer.writerows(results)

env_names = {'A': 'Env A', 'B': 'Env B'}

fig1, axes = plt.subplots(4, 2, figsize=(12, 16), sharex=True)
for idx, N in enumerate([1, 2, 3, 4]):
    for j, env in enumerate(['A', 'B']):
        ax = axes[idx, j]
        d = plot_data[N][env]
        ax.plot(tlist, d['e0'], 'k--')
        ax.plot(tlist, d['e_opt'], 'r-')
        ax.set_title(f"({chr(97 + idx*2 + j)}) $N={N}$ | {env_names[env]}")
        ax.set_ylabel(r'$\mathcal{E}(t)$')
        ax.grid(alpha=0.3)
        if idx == 3: ax.set_xlabel(r'Time $t$')
        if idx == 0 and j == 0: ax.legend()
plt.tight_layout()
plt.savefig('Fig1_Master_Dynamics.png')
plt.close()

fig2, axes = plt.subplots(3, 2, figsize=(12, 14))
delta_scan = np.linspace(0, 0.5, 15)
gamma_scan = np.linspace(0.02, 0.15, 15)

for idx, N in enumerate([2, 3, 4]):
    for j, env in enumerate(['A', 'B']):
        ax = axes[idx, j]
        map_data = np.zeros((len(gamma_scan), len(delta_scan)))
        
        for iy, g0_val in enumerate(gamma_scan):
            for ix, dlt in enumerate(delta_scan):
                e, _, _ = simulate_battery(N, env, dlt, current_gamma=g0_val)
                map_data[iy, ix] = e[-1]
                
        X, Y = np.meshgrid(delta_scan, gamma_scan)
        cp = ax.contourf(X, Y, map_data, levels=30, cmap='magma')
        fig2.colorbar(cp, ax=ax, label=r'$\mathcal{E}_{res}$')
        ax.set_title(f"({chr(97 + idx*2 + j)}) $N={N}$ | {env_names[env]}")
        ax.set_xlabel(r'Detuning $\Delta$')
        ax.set_ylabel(r'Base Decay $\gamma_0$')
        
        theo_opt = g * np.sqrt(N / (2 * gamma_scan))
        ax.plot(theo_opt, gamma_scan, 'w--', lw=2, label='Analytical $\Delta^*$')
        ax.set_xlim(0, 0.5)
        if idx==0 and j==0: ax.legend()

plt.tight_layout()
plt.savefig('Fig2_Survival_Heatmaps.png')
plt.close()

fig3 = plt.figure(figsize=(15, 10))
gs = fig3.add_gridspec(2, 2)

ax3a = fig3.add_subplot(gs[0, 0])
N_arr = np.array([1, 2, 3, 4])
d_opts_A = [plot_data[n]['A']['delta'] for n in N_arr]
ax3a.plot(N_arr, g*np.sqrt(N_arr/(2*gamma0_base)), 'k--', label=r'$\propto \sqrt{N}$')
ax3a.set_title('(a) Scaling of Optimal Detuning')
ax3a.set_xlabel('$N$')
ax3a.set_ylabel(r'$\Delta^*$')
ax3a.set_xticks(N_arr)
ax3a.legend()
ax3a.grid(alpha=0.3)

ax3b = fig3.add_subplot(gs[0, 1])
d_scan = np.linspace(0, 0.6, 15)
ergo_p, blp_p = [], []
for d in d_scan:
    e, _, b = simulate_battery(2, 'A', d)
    ergo_p.append(e[-1]); blp_p.append(b)

color = 'tab:red'
ax3b.plot(d_scan, ergo_p, 'ro-', label=r'$\mathcal{E}_{res}$')
ax3b.set_ylabel(r'$\mathcal{E}_{res}$', color=color)
ax3b.tick_params(axis='y', labelcolor=color)

ax3b_twin = ax3b.twinx()
color = 'tab:blue'
ax3b_twin.plot(d_scan, blp_p, 'bs--', label=r'BLP $\mathcal{N}$')
ax3b_twin.set_ylabel(r'BLP $\mathcal{N}$', color=color)
ax3b_twin.tick_params(axis='y', labelcolor=color)
ax3b.set_title('(b) The Non-Markovian Paradox ($N=2$)')
ax3b.set_xlabel(r'$\Delta$')
ax3b.axvline(plot_data[2]['A']['delta'], color='k', ls=':', label=r'$\Delta^*$')
ax3b.grid(alpha=0.3)

ax3c = fig3.add_subplot(gs[1, :])
adv_A = [plot_data[n]['A']['e_opt'][-1] / (n * plot_data[1]['A']['e_opt'][-1]) for n in N_arr]
adv_B = [plot_data[n]['B']['e_opt'][-1] / (n * plot_data[1]['B']['e_opt'][-1]) for n in N_arr]

ax3c.plot(N_arr, adv_A, 'bo-', label='Env A (Nuclear Decay)')
ax3c.plot(N_arr, adv_B, 'rs--', label='Env B (Thermal)')
ax3c.axhline(1.0, color='k', ls='--')
ax3c.fill_between(N_arr, 1.0, adv_A, where=(np.array(adv_A)>=1.0), color='blue', alpha=0.1)
ax3c.set_title('(c) Quantum Advantage Metric')
ax3c.set_xlabel('$N$')
ax3c.set_ylabel(r'$\mathcal{E}_N / (N \times \mathcal{E}_1)$')
ax3c.set_xticks(N_arr)
ax3c.legend()
ax3c.grid(alpha=0.3)

plt.tight_layout()
plt.savefig('Fig3_Scaling_and_Paradox.png')
plt.close()

fig4, ax4 = plt.subplots(figsize=(8, 5))
N_ext = np.arange(1, 51)
g_eff = g * np.sqrt(N_ext)
ratio = g_eff / w_b

ax4.plot(N_ext, ratio, 'k-', lw=2.5)
ax4.axhline(0.1, color='r', linestyle='--', label='RWA Breakdown Threshold (~10%)')
ax4.fill_between(N_ext, 0.1, ratio, where=(ratio >= 0.1), color='red', alpha=0.2, label='USC Regime (Model Fails)')
ax4.set_title('Regime of Validity: Breakdown of Rotating Wave Approximation')
ax4.set_xlabel('Number of Qubits ($N$)')
ax4.set_ylabel(r'Coupling Ratio $g_{eff} / \omega_b$')
ax4.grid(alpha=0.3)
ax4.legend(loc='upper left')
plt.tight_layout()
plt.savefig('Fig4_RWA_Breakdown.png')
plt.close()
#beta
from scipy.optimize import curve_fit

print("\n" + "="*50)
print("="*50)

def power_law(N, A, beta):
    return A * np.power(N, beta)

N_array = np.array([1, 2, 3, 4])

delta_fine_scan = np.linspace(0.05, 0.4, 50)
numerical_delta_opt_A = []
numerical_delta_opt_B = []

for N in N_array:
    ergo_peaks_A = []
    ergo_peaks_B = []
    for dlt in delta_fine_scan:
        e_A, _, _ = simulate_battery(N, 'A', dlt)
        e_B, _, _ = simulate_battery(N, 'B', dlt)
        ergo_peaks_A.append(e_A[-1])
        ergo_peaks_B.append(e_B[-1])
    
    numerical_delta_opt_A.append(delta_fine_scan[np.argmax(ergo_peaks_A)])
    numerical_delta_opt_B.append(delta_fine_scan[np.argmax(ergo_peaks_B)])

popt_A, _ = curve_fit(power_law, N_array, numerical_delta_opt_A)
beta_A = popt_A[1]

popt_B, _ = curve_fit(power_law, N_array, numerical_delta_opt_B)
beta_B = popt_B[1]

print(f" Analytical Beta       = 0.5000")
print(f" Empirical Beta (Env A) = {beta_A:.4f}")
print(f" Empirical Beta (Env B) = {beta_B:.4f}")
print("="*50)

print(f" Done! Execution time: {(time.time() - start_time):.2f} seconds.")
\end{lstlisting}

\bibliography{references}
\bibliographystyle{apsrev4-2}
\end{document}